 \documentclass[nohyper,12pt,letterpaper]{JHEP}
 \usepackage{epsfig}
 \newfont{\frak}{eufm10 scaled 1200}
 
 \newfont{\Bbb}{msbm10 scaled 1200} %instead of eusb10
 \newcommand{\mathbb}[1]{\mbox{\Bbb #1}}
 \DeclareSymbolFont{AMSa}{U}{msa}{m}{n}
 \DeclareSymbolFont{AMSb}{U}{msb}{m}{n}
 \let\Box\relax
 \DeclareMathSymbol{\Box}{\mathord}{AMSa}{"03}

 \def\Ref#1{(\ref{#1})}
 \def \eqn#1#2{\begin{equation}#2\label{#1}\end{equation}}

 \title{A New Approach to the Phenomenology of\\ Cosmological Supersymmetry Breaking}
 \author{T.\,Banks\\
 Department of Physics \\
 University of California, Santa Cruz, CA 95064\\
 E-mail: \email{banks@scipp.ucsc.edu}\\
 {\it and}\\
 NHETC, Rutgers University\\
 Piscataway, NJ 08540}

 \abstract{I introduce a new low energy effective description of
Cosmological SUSY breaking.  It requires the existence of a
strongly interacting gauge theory at a scale of order $10^3$ GeV,
some of whose fields carry standard model quantum numbers.  SUSY
breaking is communicated to the standard model through gaugino
masses and the Higgs multiplets. The model also provides a
possible new solution of the strong CP problem.}
 \received{} \accepted{}
 \preprint{\hepph{0408260}\\\\ \\}
 
\begin{document}
 %%%%%%%%%%%%%%%%%%%%%%%%%%%%%%%%%%%%%%%%%%%%%%%%%%%%%%%%%%%%%%%%%%%%%%%%%%%%
 % Table of contents automatic !!! %
 %%%%%%%%%%%%%%%%%%%%%%%%%%%%%%%%%%%%%%%%%%%%%%%%%%%%%%%%%%%%%%%%%%%%%%%%%%%%
 \section{\bf Introduction}

The hypothesis of Cosmological SUSY Breaking\cite{tbfolly} (CSB)
correlates the gravitino mass, $m_{3/2}$ with the cosmological
constant, according to the formula

\eqn{csb}{F_G \sim m_{3/2} M_P \sim \Lambda^{1/4}M_P .}

$\Lambda$ is viewed as a discrete, tunable parameter (perhaps
determined in the real world by galactothropic considerations),
and the {\it limiting model} with vanishing $\Lambda$ is assumed
to preserve exact $N=1,\ d=4$ super-Poincare invariance and a
discrete R-symmetry. The scaling law \Ref{csb} for the corrections
to this limit, was originally postulated on phenomenological
grounds. More recently, I provided a hand waving derivation of
this result\cite{susyhor}. The origin of the large SUSY breaking
effects is interaction with the horizon states of a stable dS
space.

The low energy effective theory resulting from such a picture is
highly constrained.  First we must find an isolated super-Poincare
invariant solution of string/M - theory.   We will call this the
{\it limiting model}. Then, small explicit R-symmetry breaking
perturbations of the low energy Lagrangian, must lead to
spontaneous breaking of SUSY.

This can only occur if the limiting model at $\Lambda = 0$ has a
massless fermion, ready to play the role of the Goldstino.   In a
previous paper\cite{susycosmopheno} I gave arguments that the
Goldstino must be a member of a vector superfield.   Those
arguments were incomplete, and the model of this paper is a
counterexample.  The models discussed in \cite{susycosmopheno}
suffered from a number of phenomenological problems.  They also
invoked a field independent Fayet-Iliopoulos term for a $U(1)$
gauge symmetry. Witten has argued\cite{ednewissues} that such a
term is inconsistent with the Dirac quantization condition.

In the present model, the Goldstino will be a chiral superfield
$G$, which transforms as a singlet under all low energy gauge
groups, and has vanishing $R$ charge.   Terms in the
superpotential, depending only on $G$, must vanish in the limiting
model.   The $G$ dependence of the superpotential for small
$\Lambda$ is generated by the mechanism described
in\cite{susyhor}.   It has the form:

\eqn{suppot}{\Lambda^{1/4} M_P^2\ w(G/M_P).}

We will arrange the rest of the dynamics of the model so that the
VEV of $|G|$ is $\ll M_P$, and consequently $F_G \sim
\Lambda^{1/4} M_P$.

\section{\bf Gaugino masses and a new low energy gauge group}

The rest of our proposal for the dynamics of the $G$ field is
motivated by the phenomenological requirement of large gaugino
masses.  We will introduce a new low energy gauge group ${\cal G}$
with a variety of chiral multiplets, some of which also transform
under the standard model gauge group.   In order to preserve
coupling unification, it is probably best to assign the standard
model quantum numbers of complete $SU(5)$ multiplets to chiral
supermultiplets transforming under ${\cal G}$.

This new gauge theory must satisfy the following criteria (we will
discuss the problem of constructing explicit examples in a later
section).

\begin{itemize}

\item It becomes strongly coupled at a scale $M_1$ but does not
break either SUSY or R-symmetry.

\item  There is a gauge invariant, renormalizable Yukawa coupling
$g_{\cal G}\int d^2 \theta G F_1 F_2$, where the $F_i $ transform
under ${\cal G}$.

\item Apart from a term $F_1 F_2$, which can be absorbed by a
linear shift of $G$, the model has no allowed relevant operators.
We choose the origin of $G$ to eliminate the relevant operator.

\item There is an R allowed coupling $g_{\mu} \int d^2 \theta GH_u H_d$.  The conventional
$\mu$ term is forbidden by a symmetry ${\cal F}$, which we will
introduce below.

\end{itemize}

When we integrate out scales above $M_1$, we obtain an effective
action for $G$.   The assumptions of unbroken R - symmetry implies
that the part of the action which depends only on $G$ consists of
a Kahler potential of the form \eqn{kahlerpot}{K = GG^* k({G\over
M_1} , {G^* \over M_1}, {{G}\over M_U},  {{ G^*}\over M_U}) , }
where $M_U$ is the unification scale.  Here we make the assumption
that the coupling of $G$ to any part of the spectrum at scales
intermediate between $M_1$ and $M_U$ is suppressed by at least an
inverse power of $M_U$.  It might even be reasonable to replace
$M_U$ with $M_P$ in this formula.   I believe that the answer to
this question depends on details of the limiting model at scales
of order $M_U$.

When combined with the superpotential \Ref{suppot}, this Kahler
potential gives a potential for $|G|$, which varies on the scale
$M_1$.   We will assume that it has a minimum at $<|G|> \sim M_1$.
SUSY is broken at this minimum. The value of the F term of G is
given approximately by

\eqn{fg}{F_G = \sqrt{K^{GG^*}(<|G|>/M_1 , <|G|>^*/ M_1)}
w^{\prime} (0) \Lambda^{1/4}M_P \sim 10^7 ({\rm GeV})^2 .}

This gives rise to a gravitino mass of order $10^{-3}$ eV.  The
formula for the mass scales with the power of $\Lambda$ predicted
in \cite{susyhor}.   The value of $w(0)$, which is a number of
order $1$, must be fine tuned to an accuracy ${\Lambda^{1/2} \over
M_P^2}$ in order to produce the correct value, $\Lambda$, for the
value of the effective potential at its minimum.  $\Lambda$ is a
fundamental input parameter in CSB, rather than a calculable low
energy effective parameter, so this fine tuning is philosophically
unexceptional.   If one wishes, one can determine the correct
value of this parameter in the real world, by applying the {\it
galactothropic principle} of Weinberg\cite{wein}, rather than
simply fitting more recent cosmological data.

To get an estimate for what we want $M_1$ to be, we calculate the
gaugino masses\footnote{This formula would follow from couplings
of the form $(G/M_1 )^a W^2$  or a Wess-Zumino coupling ${\rm ln}(
G/M_1 ) W^2$, if $<|G|> \sim M_1$. The latter form would be
natural if the ${\cal G}$ gauge theory has an accidental anomalous
$U(1)$ symmetry, $U_A$, (with anomalies coming from the standard
model), under which $G$ transforms.  For appropriate values of the
anomaly coefficients, the WZ form preserves a discrete subgroup
${\cal F}$ of this $U(1)$.  This is the symmetry we need to ensure
the naturalness of the size of the $\mu$ term. On the other hand,
if the ${\cal G}$ theory breaks $U_A$ to ${\cal F}$ by either
classical superpotential terms or a quantum anomaly, we could have
couplings of the form $(G/M_1 )^a W^2$, where $G^a$ is the lowest
dimension holomorphic ${\cal F}$ invariant, we can construct from
$G$.  If $<|G|> \sim M_1$ this gives a similar formula for gaugino
masses. The WZ form can give us a QCD axion, a possibility we
discuss below. }

\eqn{gaugino}{m_{1/2}^{(i)} \sim {\alpha_i \over\pi} \epsilon_i
{F_G \over M_1}.}

The running couplings $\alpha_i$ in this formula, are to be
evaluated at the gaugino mass scale.   $\epsilon_i$ are determined
by calculations in the strongly coupled gauge theory at scale
$M_1$. If $\epsilon_2 \sim 1$, this gives a wino mass of order
$100$ GeV if $M_1 \sim 1$ TeV.  Thus, a reasonable value for
gaugino masses requires a near coincidence between the dynamical
scale $M_1$ and the scale of CSB, $\sqrt{M_P \Lambda^{1/4}}$.

The origin of the $G$ field is determined by its coupling to the
gauge theory ${\cal G}$, so that we are not allowed to simply
absorb the conventional $\mu$ term into the VEV of $G$.  However,
if we introduce a discrete symmetry ${\cal F}$ under which $G$
transforms by a phase, and assume that the coupling $\int
d^2\theta\ G H_u H_d$ is ${\cal F}$ invariant, then the
conventional $\mu$ term will be forbidden. Like the discrete R
symmetry which guarantees Poincare invariance in the limiting
model, ${\cal F}$ will be explicitly broken by interactions with
the horizon.  This breaking is sufficiently small to ignore.  The
dominant breaking of ${\cal F}$ will come from the VEV of $G$.
${\cal F}$ is required to be a symmetry of the ${\cal G}$ gauge
theory.

\section{Baryon number, lepton number, and flavor}

A central element in CSB is the discrete R symmetry which
guarantees Poincare invariance in the the limiting model. This can
be put to other uses.   Here we will show that it can eliminate
all unwanted dimension $4$ and $5$ baryon and lepton number
violating operators in the supersymmetric standard model.   The
dimensionless coefficients of these operators will thus be
suppressed by at least ${\Lambda^{1/8} \over \sqrt{M_P}} \sim
10^{-15.5} $.   This is sufficient to account for experimental
bounds on baryon and lepton number violating processes.  The
interaction $\int d^2 \theta\ H_u^2 L^2$, will need a separate
discussion.  It should not be forbidden by R.

We will choose the R charge of SSM fields to be independent of
quark and lepton flavor, and denote it by the name of the
corresponding field. All R charges are to be understood modulo
$N$, where $Z_N$ is the R symmetry group. Flavor dependent R
charges would require many important Yukawa couplings to vanish,
and the corrections to the R symmetric limit are too small to
account for the non-zero values of these couplings.

The condition that the standard Yukawa couplings are allowed by R
symmetry is

\eqn{yukallow}{L + H_d + \bar{E} = Q + H_d + \bar{D} = Q + H_u +
\bar{U} = H_u + H_d = 2.}

Note that, although these conditions allow a term $\int d^2
\theta\ H_u H_d$, it will be forbidden by ${\cal F}$.  We will
also impose $2L + 2H_u = 2$ to allow the dimension $5$ F term
which can generate neutrino masses. The renormalizable dynamics of
the ${\cal G}$ gauge theory, must have an accidental symmetry
which forbids the generation of this term with coefficient ${1
\over M_1}$.  The combination of the accidental symmetry and the
group ${\cal F}$ introduced below, should also forbid other
dimension $5$ lepton number violating operators (both D and F
terms) with a coefficient of this scale. Neutrino masses can then
be generated by dynamics at the scale $M_U$\footnote{It is well
known that this gives neutrino masses an order of magnitude too
small to explain experiment.   We have no found no neat solution
to this problem in the present context.  Later we speculate that
this factor may be related to the small numbers which appear in
the quark mass matrix, and might be explained by the
Froggatt-Nielsen mechanism.}

Dimension $4$ baryon and lepton number violating operators in the
superpotential will be forbidden in the limiting model by the
inequalities

\eqn{nobla}{2 L + \bar{E} \neq 2} \eqn{noblb}{2 \bar{D} + \bar{U}
\neq 2, } \eqn{noble}{L + Q + \bar{E} \neq 2 .} Absence of
dimension $5$ baryon number violating operators requires
\eqn{noblc}{3Q + L \neq 2}\eqn{nobld}{3Q + H_d \neq
2}\eqn{noblf}{\bar{E} + 2 \bar{U} + \bar{D} \neq 2 ,}

The condition that there be no baryon number violating dimension
$5$ D-terms is that none of $ Q + \bar{U} - L ; $ or $U + E - D$,
vanishes.

These equations can be simplified by solving the equalities for
$\bar{E}$, $\bar{U}$, $\bar{D}$, and $H_u$ in terms of $Q,L,$ and
$H_d$.     The conditions then become

\eqn{noblei}{2 \neq 3Q + H_d \neq 3} \eqn{noblej}{ 2 Q + H_d \neq
5, }. Recall that all of these conditions are to be understood
modulo $N\geq 3$, where $Z_N$ is the discrete $R$ symmetry.

A possible solution of all of these constraints is $H_d = 1$, $Q =
0$, $L = 0$, $H_u = 1$, $\bar{E} = 1$, $\bar{D} = 1$, $\bar{U} = 1
$.   The discrete group can be $Z_{N}$ with $N \geq 3$.

Thus, if the discrete R symmetry group and its representations in
the SSM are chosen appropriately we can understand both the
absence of unacceptable baryon and lepton number violating
operators, and the presence of neutrino masses.   The observed
size of neutrino masses puts a constraint on the new physics at
the scale $M_1$ .   There must be an accidental symmetry of the
combined ${\cal G} \times SU(3) \times SU(2) \times U(1)$ gauge
theory, which forbids the lepton number violating dimension five
operators that could lead to neutrino masses of order ${<H_U>^2
\over M_1}$. The exact symmetries $R$ and ${\cal F}$ must permit
this dimension $5$ operator. They are broken by effects on the
horizon, but these mechanisms would induce this operator with a
coefficient much too small to account for neutrino masses. We
expect neutrino masses to be determined by physics near the GUT
scale, which generates this operator with coefficient $\sim
{10\over M_U}$.

\subsection{Flavor}

The simplest solution of flavor problems in this model is to
assume that the origin of flavor breaking is in physics near the
GUT scale.   The low energy theory at scales $\sim M_1$ is a
${\cal G} \otimes SU(3,2,1) $ gauge theory.  It has a large flavor
symmetry acting on quarks and leptons, which is broken only by the
standard Yukawa couplings of $H_u$ and $H_d$.   If all other
physical excitations have masses near the unification scale, the
GIM mechanism is operative and flavor changing processes occur at
acceptable levels.

In a more ambitious model, the Goldstino field $G$ might also
allow us to implement the Froggatt-Nielsen (FN) \cite{fn} mechanism
for explaining the flavor structure of quark and lepton masses and
mixing angles. The basic idea is very simple.  We postulate that
the discrete symmetry group ${\cal F}$, which commutes with the
supercharges, and under which $G$ transforms by a phase, also acts
on quarks and leptons in a flavor dependent way\footnote{It must
act on the Higgs fields in order to explain the value of the $\mu$
term. It must then act on quarks and leptons if the standard
Yukawa couplings are to preserve ${\cal F}$.  However, we have a
choice of whether this action is flavor blind or not.}. This
symmetry is of course broken for finite values of the cosmological
constant, but this explicit breaking is much smaller than the
spontaneous breaking due to the VEV of $G$. In this version of the
model, the ${\cal F}$ charges are family quantum numbers which
distinguish quark and lepton flavors.

In order to account for quark and lepton mass matrices of
appropriate size, we need, at the level of the effective field
theory containing $G$ and the standard model, nonrenormalizable
couplings between $G$ and standard model chiral multiplets, which
are scaled by $M_1$ rather than the Planck mass or unification
scale.   The non-renormalization theorem for superpotentials, and
our assumption of unbroken R - symmetry of the ${\cal G}$
dynamics, will require us to have Yukawa couplings between
standard model chiral fields, and the fields which are charged
under ${\cal G}$.  These terms could have the form $\int d^2
\theta\ T \bar{T}^A \bar{5}_A$, where $T$ is in some ${\cal G}$
representation, and a standard model singlet. $\bar{T}^A$
transforms in the conjugate representation of ${\cal G}$ and the
$[5]$ of $SU(5)$, while $\bar{5}_A$ are the usual standard model
fields which fit into the $[\bar{5}]$ of $SU(5)$.   Of course, at
this low energy level, $SU(5)$ is broken, and there is no reason
for these Yukawa couplings to satisfy $SU(5)$ relations (though we
have only written the $SU(5)$ invariant coupling explicitly).

When we integrate out physics at the scale $M_1$,  these couplings
will give rise to irrelevant couplings in the effective theory
which describes $G$ and the MSSM.  In particular, there will be
terms in the superpotential of the form $\int d^2 \theta\
\lambda_u^{ij} (G/M_1) H_u Q_i \bar{U}_j $ with similar terms for
down quarks and leptons. We assume that the ${\cal G}$ dynamics
does not break either ${\cal F}$ or the R-symmetry,
so\footnote{apart from tiny terms coming from interaction with the
horizon,} these matrix valued functions of $G/M_1$ must respect
these symmetries.   In particular, a given power of $G$ can only
appear if its ${\cal F}$ quantum numbers are neutralized by those
of the quarks or leptons. Assuming that $<G>/M_1 \sim
.2$\footnote{If we have the WZ coupling of $G$ to the standard
model gauge fields, which would follow from an unbroken accidental
$U_A$ symmetry of ${\cal G}$ physics, this assumption also
increases our estimate of gaugino masses by a factor of $5$, thus
raising the scale $M_1$.  If, on the other hand, we have couplings
of the form $G^a W^2$, then the small VEV of $G$ appears in the
numerator of our estimate for the gaugino masses.   It is no
longer possible to have a viable effective field theory.   Thus,
we can only have a FN mechanism based on $G$, if the phase of $G$
is a QCD axion.  We will see below that this is probably ruled out
experimentally.}, we get a Yukawa coupling matrix whose entries
are powers of this small parameter. This is the Froggatt-Nielsen
mechanism. There are a large number of papers on the
Froggat-Nielsen mechanism\cite{lns}, and it is well known that it
is possible to construct models of this type which give correct
predictions for the quark and lepton masses and the CKM angles.

However, Y. Nir\cite{ynir} has informed me that it is very
difficult to make phenomenologically consistent models of this
type at low scales.   Thus, this superficially attractive
possibility will probably lead to phenomenological problems. It is
likely then that the theory of flavor, like that of neutrino
masses, is associated with scales of order the unification scale.
Note by the way that the Froggatt-Nielsen mechanism for explaining
flavor hierarchies, requires two closely spaced energy scales in
order to account for the small parameter whose powers govern the
texture of the quark and lepton mass matrices.  It is interesting
to speculate that it is a power of the same small parameter which
accounts for the otherwise mysterious order of magnitude
discrepancy between the neutrino see-saw scale and the unification
scale. That is, the neutrino see-saw scale might be $\theta^b
M_U$, with $b = 1$ or $2$ and $\theta$ the Cabibbo angle.

\section{\bf $SU(2) \otimes U(1)$ breaking}

There is no reason for either of the Yukawa couplings  $g_H \int
d^2 \theta G H_u H_d$ or $g_{\cal G} \int d^2 \theta G F_1 F_2 $
to be particularly small. If we integrate out degrees of freedom
above the scale $M_1$, the first of these couplings produces
Kahler potential terms of the form ${1\over M_1} G G^* H_u H^*_u$
and ${1\over M_1} G G^* H_d H^*_d$. These terms are of order
${{g_H^4}\over 4\pi^2 }$. There are then strong ${\cal G}$
corrections to this suppressed by a further factor of ${{g_{\cal
G}}^2 \over 4 \pi^2}$.   If the couplings are not small these
terms produce significant contributions to the quadratic term in
the Higgs potential, when we insert the $F$ term of $G$. The sign
of the quadratic terms may depend on the details of strongly
interacting physics at scale $M_1$, if $g_{\cal G}$ is not small.
If the sign is negative and dominates other contributions to the
potential, we find the required breakdown of the weak interaction
gauge symmetry.  Of course, we can also get a contribution of the
same type, from the top quark Yukawa coupling, as in gauge
mediation.

Notice that $F_G$ produces tree level SUSY breaking in the Higgs
supermultiplets.  Higgs loops will then give the dominant
contribution to the splitting between the top quark and top squark
(and perhaps the bottom quark/squark splitting as well if $m_t
/m_b$ is attributed to large ${\rm tan}\beta$.  ).  Other squarks
and sleptons will get their masses predominantly through gauge
loops, as in gauge mediated models.

Thus, we can expect the spectrum of gauginos, sleptons and light
squarks to resemble that of gauge mediated models, while squark
partners of the heavy quarks, and particles in the Higgs
multiplets will have masses which depend on the new Yukawa
couplings $g_H$ and $g_{\cal G}$, and, if the latter is strong
enough, also on the details of strong ${\cal G}$ dynamics.   More
work is necessary to determine whether there are any potential
problems with existing measurements, and to sharpen the
predictions of the model for physics accessible to the LHC.

\subsection{Dark matter}

The gravitino is the lightest fermion in this model, and will be
stable.  It is relatively strongly coupled to the rest of the
system, through its Goldstino component $G$.  Thus, there will not
be a WIMP LSP candidate for dark matter.   The most likely dark
matter candidate is a cosmologically stable ${\cal G}$ hadron. The
${\cal G}$ theory is required to have a variety of exact and
accidental symmetries to account for the scale of the $\mu$
parameter, neutrino masses, {\it etc.}  It would not be surprising
to find that these implied a quasi-stable particle whose mass and
annihilation cross section were related to the scale $M_1$. More
detailed analysis will be required to determine if such a particle
is a viable dark matter candidate, but it is in the right
ballpark\footnote{Note that the mass is a bit heavier than
conventional WIMPs, but the annihilation cross section is likely
to be larger because it does not contain weak dimensionless
couplings.}

\subsection{A QCD axion and an alternative solution to the strong CP problem}

The phase of $G$ is an angular variable, which might couple to QCD
like a Peccei-Quinn-Weinberg-Wilczek axion, if $U_A$ symmetry is
unbroken by ${\cal G}$ dynamics. However, the axion decay constant
is in a range which is almost certainly ruled out by experiment.
In our low energy model, at scale $M_1$ we can postulate an
accidental axial symmetry which acts on $G$, and guarantees that
its Kahler potential has the form $K(G \bar{G} / M_1^2 )$, and
that the couplings to standard model gauge bosons have the WZ
form. In this approximation, $G$ is a QCD axion with decay
constant $\sim M_1$. Dynamics at and above the unification scale
has no reason to preserve this symmetry.  It need only preserve
the discrete subgroup ${\cal F}$. Assume that this group is $Z_p$,
and that $G$ has charge $q$. Then there is some lowest power $G^a$
($a \geq 2$) which is invariant, and there can be terms in the
Kahler potential of the form \eqn{deltak}{\delta K = {{G^{a + 1}
\bar{G} + h.c. } \over M^a },} where $M$ might be the unification
scale or the Planck scale. This will give a high energy
contribution to the would be axion mass\footnote{There can be
other contributions coming from the superpotential, but these give
smaller contributions to $m_a$.}\eqn{mahigh}{(m_a^{high})^2 \sim
{M_P \Lambda^{1/4} \over M_1} ({M_1 \over M})^{a/2}.} The QCD
contribution is \eqn{maqcd}{m_a^{QCD} \sim {{(100 {\rm
MeV})^2}\over M_1}}

For $M_1 \sim 1$ TeV the ratio of the high energy contribution to
the QCD contribution is  $\sim 10^9 ({M_1 \over M})^{a/2}$.  Even
for $a=2$ and $M \sim 10^{15} $ GeV, the QCD contribution
dominates.    So we appear to have an axion but in an
experimentally forbidden range.

There are three ways out of this problem.   The simplest is to
assume that only ${\cal F}$ and not the full $U_A$ group is a
classical symmetry of the ${\cal G}$ Lagrangian.  That is, other
terms in the superpotential for chiral fields charged under ${\cal
G}$ break $U_A$ down to a discrete subgroup containing ${\cal F}$.
Alternatively, the classical $U_A$ symmetry could be broken by a
${\cal G}$ anomaly.  In either case, strong dynamics at the scale
$M_1$ would give both scalar components of $G$ masses of order
${{M_P \Lambda^{1/4}}\over M_1}$.

A more interesting possibility is that $U_A$ is a symmetry, but
that our visible QCD axion actually evades the conventional
experimental bounds.   At the effective Lagrangian level it
appears that the axion to goldstino pair amplitude is larger by a
factor of $({\alpha_i \over \pi})^{-1}$ (where $\alpha_i$ is a
standard model gauge coupling) than any decay into visible
products.  If the visible branching ratios are small enough, the
conventional bounds might be evaded.   A note of caution here is
that the matrix element of the leading operator mediating the
decays into goldstinos may be chirally suppressed on
shell\footnote{I would like to thank S. Thomas for pointing this
out to me.}.  Non-leading operators would contribute to p-wave
decays into goldstinos and would have a suppression factor of
order $(m_a / M_1)^2$ relative to the estimate above.  This more
than makes up for the gauge coupling suppression of visible
decays. If this is indeed the case, the visible decays would
dominate and such an axion is ruled out.  We would then have to
invoke breaking of $U_A$ at the scale $M_1$ to construct a viable
model.

In fact, our model {\it may} contain an alternative solution of
the strong CP problem.  To obtain it, we must make {\it another}
assumption about the elusive ${\cal G}$ gauge theory: it should
have automatic CP conservation.  That is, the exact discrete R and
${\cal F}$ symmetries, the ${\cal G}$ gauge symmetry and
renormalizability should guarantee the existence of an accidental
CP symmetry of the ${\cal G}$ Lagrangian, under which $G$ goes
into its complex conjugate. In particular, in order to shift away
the topological term in the ${\cal G}$ gauge Lagrangian, we rotate
by the $U(1)$ R transformation under which all gauginos rotate,
and all chiral superfields have R-charge 0. If the chiral
multiplets in the theory fall into $K$ irreducible ${\cal G}$
multiplets, the gauge interactions in the model are invariant
under $K-1\ {\cal G}$-anomaly free $U(1)$ symmetries, which are
linear combinations of the phase rotations of the individual
multiplets.  If there are several multiplets in the same ${\cal
G}$ representation, they are also invariant under ${\cal
G}$-anomaly free $SU(m)$ transformations.   The standard model
gauge group is a subgroup of this anomaly free group, which also
leaves all Yukawa couplings invariant\footnote{Recall our
assumption that the model contains no relevant operators allowed
by the symmetries.}.   The condition for CP invariance is that the
full anomaly free group can transform away all phases in the
Yukawa couplings that are allowed by ${\cal G} \otimes SU(3,2,1)
\otimes R \otimes {\cal F}$.

One consequence of this assumption for the effective theory below
$M_1$ is that the Kahler potential $G\bar{G}K(G/M_1, G^* /M_1 )$
is CP invariant.   We will also make the technically natural
assumption that the minimum of the potential, which is derived
from $K$ once we add the superpotential to the limiting
Lagrangian, is CP conserving; {\it i.e.} $<G>$ is real.

Furthermore, CP invariance of the strong ${\cal G}$ dynamics
guarantees that the coefficients $\epsilon_i$ in the couplings
$\epsilon_i (G/M_1)^a W_i^2$ to the standard model gauge bosons,
are all real, and do not shift the value of $\theta_{QCD}$ when
the VEV of $G$ is turned on.

 Now consider the low energy Lagrangian for the standard model
 coupled to $G$, still in the limiting model with $\Lambda =0$.
 We can use the a combination of the $U(1)_R$ of gaugino rotation,
 the $U(1)_A$ which rotates all quark and anti-quark
 superfields by the same phase, and an
 equal phase rotation of the Higgs superfields\footnote{The
 independent linear combination of phase rotations on $H_u$ and
 $H_d$ is the gauged weak hypercharge.  We will always choose a
 gauge in which the VEVs of the two Higgs fields have equal
 phase.}to eliminate both $\theta_{QCD}$ and
 ${\rm arg} {\rm det} g_u g_d$, (where $g_{u,d}$ are the up and
 down quark Yukawa coupling matrices), and to make the phase of
 the coupling $g_{\mu}$ in $g_{\mu} \int d^2 \theta G H_u H_d$
 real and negative.

 In a nutshell, what we have shown is that, like the standard
 model before the discovery of the $U(1)_A$ anomaly, the limiting
 $\Lambda = 0$ model has all CP violation concentrated in the
 usual Jarlskog parameter of the CKM matrix.

 Now consider what happens when $\Lambda \neq 0$.   The
 superpotential for $G$ comes from Planck scale physics near the
 horizon.  It has no apparent reason to be CP invariant.   In
 particular, the coefficent $w^{\prime} (0)$ which determines
 $F_G$ might be complex.  Write $w^{\prime} (0) = |w^{\prime} (0)
 |e^{ia}$.  In the low energy effective Lagrangian, below the
 scale $M_1$,  $F_G$ appears linearly and quadratically.  The
 quadratic terms have the form $F_G F_G^*$ .   Thus, the phase $a$
 appears only in the gaugino masses
 \eqn{mgaugino}{m_{1/2}^i \tilde{g}^i
 \tilde{g}^i = r^i {|F_G|\over M_1} e^{i a} \tilde{g}^i
 \tilde{g}^i} ($r^i $ are real numbers of order ${\alpha_i \over\pi}$)
 and in the ``b - term"
 \eqn{bterm}{m_{ud}^2 h_u h_d = g_{\mu} |F_G| e^{i a} h_u h_d + c.c.}
The latter term is the only term in the effective potential that
depends on the overall phase $e^{i (a_u + a_d)}$ of the Higgs
fields.  It is minimized by $e^{i (a_u + a_d)} = e^{- ia}$.

When the Higgs VEVs are substituted in the Yukawa couplings, this
generates a phase for the determinant of the quark mass matrix
\eqn{massmat}{{\rm arg\ det} {\cal M} = - 3ia }  We can eliminate
both the phase of the gluino mass and that of the quark
determinant by doing a $U(1)_R$ rotation with angle satisfying
$e^{2 i\theta_R} = e^{- i a}$, and a $U(1)_A$ rotation with $e^{12
i \theta_A} = e^{3 i a}$. A particular solution of these equations
is \eqn{thetaeqn}{\theta_R = - 2\theta_A = - {a\over 2}} Now
recall that the Dynkin index, which determines the $SU(3)$ anomaly
of the $U(1)$ rotation of a single Weyl fermion is $3$ for the
adjoint of $SU(3)$ and ${1\over 2}$ for the fundamental. $U(1)_A$
combines the $U(1)$ rotations of $12$ Weyl fermions so it shifts
$\theta_{QCD}$ by twice as much as a rotation by the same angle in
$U(1)_R$.   {\it But we have found that the $U(1)_A$ rotation we
need to eliminate the phase of the quark determinant is half as
large and has the opposite sign of the rotation we need to
eliminate the gluino mass !}   Thus, there is is no net
$\theta_{QCD}$.

There are several issues which must be checked before concluding
that this is a real solution of the strong CP problem.   We have
only examined the effect of the superpotential $w(G)$ on the low
energy effective theory.   In fact, this interaction exists in the
effective theory at any scale below the Planck scale.  It is not
clear whether renormalization effects coming from this term in the
Lagrangian above the scale $M_1$ can invalidate our argument.

We have also used $U(1)_R$ transformations, without regard to
their effect on irrelevant perturbations of the low energy theory.
In our first use of a $U(1)_R$, to analyze the strong ${\cal G}$
dynamics, these corrections would be proportional to powers of
${M_1 \over M_U} \leq 10^{-11}$ and do not effect the argument.
However we used a second $U(1)_R$ transformation below the scale
$M_1$.   Here the irrelevant operators are scaled by $1/M_1$,
probably multiplied by powers of $\alpha_i / \pi$ from the
standard model gauge interactions above $M_1$.   These could
provide new sources of CP violation and one must check that they
do not induce a neutron electric dipole moment which contradicts
experiment.

We have given an argument to the effect that the value of
$\theta_{QCD}$ at the weak scale vanishes in our model.  One must
also check that the renormalization of this parameter between the
weak scale and the scale at which the neutron EDM is measured, is
small.  A general argument to this effect, valid for a large class
of theories, was given in \cite{tbynns}.  One must check that that
argument applies to the present model.

Finally, there is the issue of whether the many constraints on the
${\cal G}$ theory are such that they force us to have very light
or massless G-hadrons, which contradict experimental bounds.  In
the next section we will note that such hadrons could even spoil
our mechanism for SUSY breaking, which would mean that our model
could not actually be a low energy effective Lagrangian for CSB.
As far as I can see, the constraint of solving the strong CP
problem does not add to this worry, but in the absence of a
specific model, it is hard to be certain.

\section{In search of a microscopic model}

With the exception of the remarks about dark matter in the
penultimate subsection, the phenomenological properties of our
model can be expressed in terms of a lagrangian involving only the
fields of the SUSic Standard Model and the Goldstino field, $G$.
That Lagrangian is however non-renormalizable, and several of its
crucial properties ({\it e.g.} the sign of the quartic term in the
Kahler potential) depend on physics at the scale $M_1$ and above.
Furthermore, at least one of the scalar components of the $G$
field gets a mass close to $M_1$ once R-symmetry breaking is taken
into account. One feels a moral compulsion to present a UV
completion of the model which is valid up to the unification
scale, and explains the details of the low energy effective
lagrangian.  So far, I have not come up with such a model.

The obvious candidate for the ${\cal G}$ theory is SUSY QCD with
$N_F = N_C + 1$.   This model is asymptotically free, and has a
vacuum state preserving both SUSY and a chiral R symmetry.   We
can couple $G$ to combinations of the gauge invariant operators
$\bar{F}_b F^a$. If $N_F = 5$ it has an anomaly free $SU(5)$
flavor symmetry, into which we can embed the standard model. The
Yukawa couplings of $G$ should of course preserve the standard
model gauge symmetry.

Unfortunately, this model does not break SUSY\footnote{I would
like to thank N. Seiberg for helping me to analyze this system.},
even when supplemented by the superpotential $w(G)$. The
superpotential depends on the composite degrees of freedom $M_b^a
= \bar{F}_b F^a$ and the baryons $B_a = \epsilon_{a, a_1 \ldots
a_4} F^{a_1}_{A_1} \ldots F^{a_4}_{A_4} \epsilon^{A_1 \ldots
A_4}$, and $\bar{B}^a = \epsilon^{a, a_1 \ldots a_4}
\bar{F}_{a_1}^{A_1} \ldots \bar{F}_{a_4}^{A_4} \epsilon_{A_1
\ldots A_4}$.   For small $G$ it has the form \eqn{suppotqcd}{W =
F_G G + M_1 G Y_a^b M_b^a + y_B B_a \bar{B}^b M_b^a + y_D {\rm
det} M ,} where $F_G$ was defined above.   This superpotential has
SUSic minima, a single point with $B = \bar{B} = 0$, and a
baryonic branch.

The model possesses several features whose generality is to be
feared.   Most asymptotically free gauge theories which can couple
to the standard model will have continuous chiral symmetries with
standard model anomalies.   Indeed, anomaly matching is likely to
be a key argument in showing that the model does not break R
symmetry.  This means that the model predicts additional massless
degrees of freedom.  In particular, the operator to which $G$
couples may often be a free massless field $M$ at low energies, in
the theory without the coupling to $G$. The $G$ coupling then
provides a mass term and the expectation value of the free field
$M$ can cancel the $F$ term that comes from interactions with the
horizon. It is thus a significant challenge to produce a
microscopic model which accomplishes our goals.   On the other
hand, if these arguments do not lead to a no-go theorem, we can
hope that the low energy dynamics will be highly constrained.

\section{\bf Conclusions}

I have presented an effective field theory of SUSY breaking which
is based on the the idea of CSB.  At the effective level, it
contains one new singlet chiral superfield $G$.  When the c.c. is
set to zero, the theory is exactly super-Poincare invariant.  It
is also invariant under a discrete complex R symmetry as well as
an ordinary discrete symmetry ${\cal F}$.   The R charge of G is
zero, and if there are no low energy fields with R charge $2$,
this guarantees that $G$ is massless.  R symmetry charges of
standard model fields are chosen to allow all of the SSM couplings
while forbidding all dimension $4$ and $5$ operators that violate
baryon and lepton number, except for dimension $5$ operators that
generate neutrino masses.

$G$ has a renormalizable $g_H \int d^2 \theta  G H_u H_d$ coupling
to the unique gauge invariant dimension $2$ operator in the SSM.
${\cal F}$ symmetry charges are chosen to ensure that the $\mu$
term of the SSM can be generated only by the VEV of
$G$\footnote{At this level of analysis, we could simply absorb the
$\mu$ term in $G$, but when terms in the ${\cal G}$ lagrangian are
taken into account, we must impose a symmetry to eliminate it.}.

SUSY is broken, and the VEV of $G$ determined, by a superpotential
of the form $M_P^2 \Lambda^{1/4} w(G/M_P)$, combined with a
non-renormalizable Kahler potential $K = G G^* k(G/M_1 , G^*/M_1
)$, with $M_1 \ll M_P$. The form of the superpotential follows
from the hypothesis of CSB.   The constraint on the magnitude of
$M_1 $ is chosen for blatantly phenomenological reasons.  There is
then a SUSY breaking minimum with $ |G| \sim M_1 $. The constant
term in the expansion of $w$ around $G = 0$ can be used to tune
the effective cosmological const to its observed value.  The VEV
of $G$ gives rise to a $\mu$ term for the SSM, while the VEV of
$F_G \sim M_P \Lambda^{1/4}$ gives rise to the $\mu_{ud}^2$ scalar
mass $h_u h_d + h.c.$ .

Gaugino masses are generated by couplings $ \epsilon_i {\alpha_i
\over \pi} (G/M_1)^a W_{\alpha}^2 $\footnote{We can also have
couplings of the form ${\rm ln} (G/M_1 ) W_{\alpha}^2$ if $U_A$
symmetry is unbroken by strong dynamics at the scale $M_1$.  This
leads to a QCD axion, which is probably ruled out by experiment.}.
In order that the various low energy SUSY breaking parameters be
within experimental bounds, we must choose $M_1 \sim 1 $ TeV. This
means that we are straining the bounds of effective field theory,
and it behooves us to construct a more microscopic model of
physics at the scale $M_1$. This has proved to be difficult. Let
me summarize the constraints on the microscopic theory.

\begin{itemize}

\item   It must generate the dynamical scale $M_1$.

\item  It must not break SUSY, $R$ or ${\cal F}$, explicitly or
spontaneously.

\item  It must have a marginal coupling $g_{\cal G}\int d^2  \theta G O_2$, where $O_2$ is
a dimension two operator of R charge two, which is consistent with
all of the symmetries of the problem.  The non-trivial Kahler
potential for $G$, which fixes $<|G|> \sim M_1$, is generated via
this coupling. Note that we need the ${\cal F}$ symmetry to
explain either the absence of the SSM $\mu$ term, or a term
involving $O_2$ without $G$ (either one of these could be shifted
into the VEV of $G$). It is important that the operator $O_2$ not
appear as a dimension one field in the effective theory below
$M_1$.  This is the property that we have found hard to realize in
explicit models.

\item  It might have an accidental $U(1)$ symmetry, $U_A$, with standard
model anomalies, which can explain a logarithmic form of the
non-renormalizable couplings of $G$ to the standard model gauge
fields. ${\cal F}$ is an anomaly free discrete subgroup of $U_A$.
If this accidental symmetry is unbroken by either classical or
quantum effects at the scale $M_1$ the model has a QCD axion. The
axion decay constant is in a range that is probably ruled out by
experiment, unless the axion decay mode into gravitinos dominates
its visible decays.

\item We have found a tentative solution to the strong CP problem
without an axion, if the strong ${\cal G}$ dynamics is
automatically CP conserving.

\item

We also need an accidental symmetry of the theory at scale $M_1$,
which will prevent the generation of the dimension five operator
$\int d^2 \theta H_u^2 LL$ with coefficient ${1\over {M_1}}$, but
permit it with a coefficient an order of magnitude below the
unification scale.  This accidental lepton number, could
conceivably be a subgroup of $U_A$ larger than ${\cal F}$.

\end{itemize}

The model also contains tantalizing hints of connections to other
important problems in particle physics. We have mentioned the
remote possibility of a viable QCD axion. We have also noted that
the discrete symmetry ${\cal F}$ could play the role of a
horizontal symmetry, if we allow it to be generation dependent. In
order to implement this, one must assume that $|G|/M_1$ is small,
perhaps of order $.2$, the Cabibbo angle. Note that, given the
formula for gaugino masses, this could also raise the scale $M_1$
and make our effective field theory approximation a little more
palatable, but only if the $G$ coupling to standard model gauge
fields has the WZ form. This implies the low scale axion, which is
probably ruled out.  If we instead have couplings of the form $(G/
M_1)^a W^2$, a small value of $<G>$ would force the scale $M_1$ to
be very low and the model is ruled out. In addition to this
problem, conversations with Y. Nir have convinced me that
implementation of the idea that ${\cal F}$ is a horizontal
symmetry is likely to remove the attractive flavor properties of
the model. Indeed, as it stands, the only terms in the lagrangian
which break the large flavor symmetry of the standard model gauge
theory, are the Yukawa couplings to the Higgs boson. There is a
consistent picture in which all issues having to do with flavor
and neutrino masses, are fixed at scales within an order of
magnitude below the GUT scale. The model then contains a natural
GIM mechanism and there is no SUSY flavor problem.

Both of these issues should be explored more thoroughly.
Additional work is also necessary to determine whether the
mechanism for $SU(2) \times U(1)$ breaking in this model requires
tuning.  There are extra contributions to the quadratic term in
the Higgs potential (beyond those familiar from gauge mediation),
which involve the strong coupling dynamics of the theory at scale
$M_1$.   We may have to wait for an explicit model of this sector
before we can assess the answer to this question.

Some readers may be disturbed by the near coincidence between the
scale $M_1$ and the logically independent scale $\sqrt{M_P
\Lambda^{1/4}}$.  We must postulate this coincidence for
phenomenological reasons, but there is no apparent dynamical
reason for it in the low energy model.   I would love to find such
a dynamical mechanism, but I am not sure that the coincidence is
so much worse than that between the weak scale and the QCD scale
(which some authors have found puzzling enough to require an
explanation).   According to the tenets of CSB, R symmetric
couplings, like the ${\cal G}$ gauge coupling at the unification
scale, are determined (up to very small corrections) by their
values in the limiting SUSic theory.  If the limiting SUSic model
is unique, and the value of this coupling (which determines $M_1$)
happened to work out correctly, we would consider it a great
triumph.   So, the puzzle of the coincidence between $M_1$ and the
SUSY breaking scale may be, like flavor and neutrino masses, a
puzzle that will only be resolved when we learn the full high
energy theory.

To summarize this summary,  I have found what appears to be an
attractive model of SUSY breaking, motivated by the ideas of CSB.
It solves many of the problems of other approaches, and presents
us with a new candidate solution of the strong CP problem. More
work is necessary to completely assess its phenomenological
viability, and to work out its detailed predictions for physics at
the TeV scale.

\section{Acknowledgments}

I would like to thank M.Dine, W. Fischler, E.Gorbatov, A.Nelson,
Y.Nir, P.J.Fox, N.Seiberg, and S.Thomas for conversations which
contributed to this work.

This research was supported in part by DOE grant number
DE-FG03-92ER40689.

%%%%%%%%%%%%%%%%%%%%%%%%%%%%%%%%%%%%%%%%%%%%%%%%%%%%%%%%%%%%%%%%%%%%%%%%%=

%%%

%                      REFERENCES                                        =

  %

%%%%%%%%%%%%%%%%%%%%%%%%%%%%%%%%%%%%%%%%%%%%%%%%%%%%%%%%%%%%%%%%%%%%%%%%%=

%%%

%\newpage


\begin{thebibliography}{19}
%here 19 is the widest mark...
%-----Type it \bibitem[how it is marked]{how we call it}Authors, hep-th/.=
%-----Citations are then made by \cite{how we call it} in text ----------=
%-----\bibitem without [how it is denoted] is numbered 1,2,3....

\bibitem{tbfolly}T.~Banks, {\it Cosmological breaking of
supersymmetry}, hep-th/0007146, Int. J. Mod. Phys. A16, 910-921,
(2001).
\bibitem{susyhor}T.~Banks, {\it Breaking susy on the horizon},
hep-th/0206117
\bibitem{susycosmopheno}T.~Banks,{\it The phenomenology of cosmological supersymmetry
breaking} , hep-ph/0203066
\bibitem{ednewissues}E.~Witten, {\it New issues in manifolds of
$SU(3)$ holonomy}, Nucl. Phys. B268, 79, (1986).
\bibitem{wein}S.~Weinberg, {\it Anthropic bound on the
cosmological constant}, Phys. Rev. Lett. 59, 2607, (1987).
\bibitem{fn}C.~Froggatt, H.B.~Nielsen, {\it Hierarchy of quark masses, cabibbo angles
and CP violation}, Nucl. Phys. B147, 277, (1979).
\bibitem{lns}M.~Leurer, Y.~Nir, N.~Seiberg, {\it Mass matrix
models}, hep-ph/9212278, Nucl. Phys. B398, 319, (1993); {\it Mass
matrix models: the sequel}, hep-ph/9310320, Nucl. Phys. B420, 468,
(1994), and references therein.
\bibitem{tbynns}T.~Banks, Y.~Nir, N.~Seiberg, {\it Missing (up)
mass, accidental anomalous symmetries, and the strong CP problem},
hep-ph/9403203, Proceedings of 2nd IFT Workshop on Yukawa
Couplings and the Origin of Mass, p.26, (1994), Gainesville, FL
11-13 Feb. 1994.
\bibitem{ynir} Y. Nir, {\it Private Communication}
\end{thebibliography}
\end{document}